\documentclass[prl,reprint,superscriptaddress,preprintnumbers]{revtex4-2}
\pdfoutput=1
\usepackage{amsmath,amssymb,graphics, comment}
\usepackage{verbatim}
\usepackage{graphicx}
\usepackage{hyperref}
\usepackage{color}
\usepackage{mathrsfs}
\usepackage{ragged2e}
\usepackage{float}

\usepackage{esint}

\usepackage{color}
\definecolor{dark-gray}{gray}{0.20}
\definecolor{gray}{gray}{0.30}
\definecolor{light-gray}{gray}{0.80}
\definecolor{dark-red}{rgb}{0.7,0,0}
\definecolor{dark-green}{rgb}{0.1,0.4,0}
\definecolor{dark-blue}{rgb}{0.3,0.3,0.7}
\definecolor{light-blue}{rgb}{0.8,0.8,1}
\definecolor{blue}{rgb}{0,0,1}
\definecolor{red}{rgb}{1,0,0}
\definecolor{green}{rgb}{0,1,0}

\newcommand {\cA}{{\cal A}}

\newcommand {\cD}{{\cal D}}

\newcommand {\cL}{{\cal L}}
\newcommand {\cM}{{\cal M}}
\newcommand {\cN}{{\cal N}}

\newcommand {\cQ}{{\cal Q}}
\newcommand {\cR}{{\cal R}}

\newcommand {\cY}{{\cal Y}}

\newcommand{\de}{{\nabla}}

\def\a{\alpha}
\def\b{\beta}

\def\d{\delta}

\def\l{\lambda}
\def\m{\mu}

\def\ri{{\rm i}}

\newcommand{\hm}{{{m}}}
\newcommand{\hn}{{{n}}}

\newcommand{\ha}{{{a}}}
\newcommand{\hb}{{{b}}}
\newcommand{\hc}{{{c}}}
\newcommand{\hd}{{{d}}}
\newcommand{\he}{{{e}}}

\newcommand{\hal}{{{\a}}}

\newcommand{\ve}{\varepsilon}

\newcommand{\pa}{\partial}
\newcommand{\hf}{\frac12}

\newcommand{\bsubeq}{\begin{subequations}}
\newcommand{\esubeq}{\end{subequations}}

\newcommand{\be}{\begin{equation}}
\newcommand{\ee}{\end{equation}}
\newcommand{\bea}{\begin{eqnarray}}
\newcommand{\eea}{\end{eqnarray}}
\newcommand{\non}{\nonumber}
\newcommand{\ba}{\begin{array}}
\newcommand{\ea}{\end{array}}
\newcommand{\nn}{\nonumber}

\begin{document}

\title{Holographic origin of $a$-maximization and higher-derivative AdS$_5$/CFT$_4$}

\author{Kiril Hristov}
\affiliation{Faculty of Physics, Sofia University ``St.\ Kliment Ohridski'', J. Bourchier Blvd. 5, 1164 Sofia, Bulgaria}
\affiliation{INRNE, Bulgarian Academy of Sciences, Tsarigradsko Chaussee 72, 1784 Sofia, Bulgaria}

\author{Saurish Khandelwal}
\affiliation{Institut f\"ur Theoretische Physik Leibniz Universit\"at Hannover
Appelstra\ss e 2, 30167 Hannover, Germany}

\author{Yi Pang}
\affiliation{Center for Joint Quantum Studies and Department of Physics, School of Science,
Tianjin University, 135 Yaguan Road, Tianjin 300350, China}
\affiliation{Peng Huanwu Center for Fundamental Theory, Hefei, Anhui 230026, China}

\author{Gabriele Tartaglino-Mazzucchelli}
\affiliation{School of Mathematics and Physics, University of Queensland,  St Lucia, Brisbane, Queensland 4072, Australia}

\begin{abstract}
We develop a consistent partially off-shell framework for evaluating higher-derivative actions of five-dimensional $\cN=1$ gauged supergravity with abelian vector multiplets on AdS$_5$. Using the superconformal formalism, we show that the resulting holographic expression reproduces the trial $a$-anomaly coefficient of the dual conformal field theory, identifying the supergravity equations of motion with $a$-maximization.  We present an exact correspondence between Chern–Simons couplings and the anomaly polynomial of the boundary theory. We illustrate our proposal  by applying it to all known two- and four-derivative actions, including the “off-diagonal’’ invariants never before considered in the gauged supergravity literature. Finally, we argue that all invariants beyond four-derivative yield no genuinely new contributions to asymptotically AdS$_5$ BPS backgrounds, but instead reduce to linear combinations of the established two- and four-derivative actions.
\end{abstract}
\date{\today}
\maketitle


\section{Introduction}
\label{sec:intro}

The AdS/CFT correspondence is one of the major developments in modern theoretical physics, providing a non-perturbative definition of certain quantum gravity theories in terms of conformal field theories and yielding powerful insights into black-hole physics, strongly coupled gauge dynamics, and even condensed-matter systems. Most of our understanding of AdS/CFT relies on the leading large-$N$ and two-derivative supergravity approximations, where genuine stringy and quantum-gravity effects are suppressed. Recent progress in exact AdS$_5$/CFT$_4$~\cite{Cremonini:2008tw,Baggio:2014hua,Bobev:2021qxx,Liu:2022sew,Bobev:2022bjm,Cassani:2022lrk,Gold:2023ymc,Cassani:2023vsa,Cassani:2024tvk,Ma:2024ynp} has highlighted the need for results beyond these approximations.  On the CFT side, major advances have been made by using supersymmetric localization; see \cite{Pestun:2007rz,Pestun:2016zxk}. On the gravity side, by contrast, a systematic understanding of higher-derivative (HD) corrections is still lacking. This is essential for performing new precision tests in holography and for probing the structure of quantum gravity.

In this work, we initiate a systematic study of HD corrections in matter-coupled gauged five-dimensional supergravity relevant to holography. The superconformal (or off-shell) formalism \footnote{By a mild abuse of terminology, we use the terms \emph{superconformal} and \emph{off-shell} interchangeably when referring to the 5d supergravity formalism.} provides a convenient framework for constructing supersymmetry-invariant terms with various derivative orders, though distinct invariants typically involve different auxiliary fields and may yield linearly dependent on-shell actions. Building on advances in superconformal tensor calculus and superspace techniques---see, e.g., \cite{Zucker:1999fn,Kugo:2000hn,Kugo:2000af,Fujita:2001kv,Bergshoeff:2001hc,Kugo:2002vc,Bergshoeff:2002qk,Zucker:2003qv,Bergshoeff:2004kh,Hanaki:2006pj,Ozkan:2013nwa,Butter:2014xxa,Ozkan:2016csy,Gold:2023dfe,Gold:2023ymc,Gold:2023ykx,Kuzenko:2007aj,Kuzenko:2007cj,Kuzenko:2007hu,Kuzenko:2008wr,Kuzenko:2008kw,Kuzenko:2014eqa} for the 5d case as well as the reviews \cite{Freedman:2012zz,Lauria:2020rhc,Kuzenko:2022skv,Kuzenko:2022ajd,Ozkan:2024euj} and references therein---we organize all invariants according to their \emph{on-shell} Chern-Simons (CS) terms. These CS couplings are dual to the ’t Hooft anomalies in the 4d SCFT,~\cite{Witten:1998qj,Henningson:1998gx,Tachikawa:2005tq,Aharony:2007dj}, with the structure of the CS terms holographically encoding the anomaly coefficients. This underpins the perturbative behavior of all supersymmetric partition functions of the dual theory.

\begin{table}[ht]
	\begin{center}
	\setlength{\tabcolsep}{6pt}
	\begin{tabular}{ c || c| c | c | } \hline
	\multicolumn{1}{|c||}{\textbf{invariant}} & $k_{IJK}$ & $k_I$  & Ref.  \\ \hline \hline
	\multicolumn{1}{|c||}{$2 \partial$ \text{theory}} & $3\, C_{I J K}$ & 0 & \cite{Zucker:1999ej,Zucker:1999fn,Zucker:2003qv,Kugo:2000hn,Kugo:2000af,Fujita:2001kv,Kugo:2002vc,Bergshoeff:2001hc,Bergshoeff:2002qk,Bergshoeff:2004kh,Butter:2014xxa}  \\ \hline
    \multicolumn{1}{|c||}{$\cR\, F$} & $-12\, g_{(I}d_{JK)L}a^L$ & $0$ & \cite{Ozkan:2016csy,Hristov:2026}\\
     \hline \hline
	\multicolumn{1}{|c||}{\text{Weyl}$^2$} & $\, 8\, \alpha_{( I} g_J g_{K )}$ & $\alpha_I$ & \cite{Hanaki:2006pj, Gold:2023ymc} \\ \hline
	\multicolumn{1}{|c||}{$\log$} & $-4\, \beta_{( I} g_J g_{K )}$ & $0$ & \cite{Gold:2023ykx, Gold:2023ymc} \\ \hline
	\multicolumn{1}{|c||}{$\cR^2$} & ${12}\, \gamma_{( I} g_J g_{K )}$ & $0$ & \cite{Ozkan:2013nwa, Gold:2023ykx}  \\ \hline\hline
	\multicolumn{1}{|c||}{$\cR\, F^3$} & $-12\, \eta \, g_{( I} g_J g_{K )}$ & $0$ & \cite{Ozkan:2016csy,Hristov:2026}\\ \hline
	\multicolumn{1}{|c||}{$F^4$} & $0$ & $0$ & \cite{Ozkan:2016csy,Hristov:2026} \\ \hline
 	\end{tabular}
	\end{center}
    \vspace{-0.21cm}
	\caption{The contribution to the Chern-Simons coefficients in \eqref{eq:on-shellCS} from different off-shell invariants. $C_{IJK}, g_I, \alpha_I, \beta_I, \gamma_I, \eta, a^{I}, d_{IJK}$  denote the various coupling constants of the corresponding supersymmetric actions, discussed in due course. The $g_I$'s play the role of the Fayet-Iliopoulos parameters that decide which linear combination of abelian vectors is identified as $U(1)_R$ symmetry.}
	\label{tab:solutions}
\end{table}
    \vspace{-0.15cm}

Concretely, after integrating out the needed auxiliary fields~\footnote{Here, ``auxiliary’’ refers to fields without kinetic terms in the two-derivative theory.}, only two types of CS terms appear in the supergravity Lagrangian
\be
\label{eq:on-shellCS}
	\cL_\text{CS}= \frac1{24} \left( k_{IJK}\, v^I \wedge {\rm d} v^J \wedge {\rm d} v^K -3\, k_I\, v^I \wedge \cR_{ab} \wedge \cR^{ab} \right)\ ,
\ee
where $v^I$ are the abelian gauge fields in the vector multiplets, and $\cR_{ab}$ is the Riemann curvature two-form. The coefficients $k_{I J K}$ and $k_I$ will be directly related to the cubic and linear 't Hooft anomalies of the dual SCFT, respectively. These on-shell CS terms, involving only physical fields, can originate from different off-shell invariants in 5d gauged supergravity, where we have integrated out only a small subset of all auxiliary fields and kept the rest of the action off-shell. Thus our partially off-shell approach circumvents the complexity of solving all auxiliary equations of motion while retaining direct holographic relevance.

Guided by the CS-term classification, we consider the AdS$_5$ background with radius $l$, written in global coordinates with $\mathbb{R}\times$S$^3$ boundary. The conditions imposed by maximal supersymmetry uniquely determine all background fields except for the vector-multiplet scalars $W^I$, which remain unspecified constants. Evaluating the supergravity action on this background using the minimal holographic renormalization scheme—without imposing the remaining scalar equations of motion— yields
\be
\label{eq:AdSoff}
	I_{AdS_5}^\text{off-shell} (W)  = \frac{\pi^2\, \beta}{4}\, \big(l^2\, k_{IJK}\, W^I W^J W^K - 18\, k_I\, W^I  \big)\ ,
\ee
where the (Wick-rotated) time direction is periodically identified with period $\beta$. The resulting expression depends only on the effective CS coefficients, not on the detailed couplings of the underlying off-shell invariants. The fully on-shell AdS$_5$ action follows from extremizing \eqref{eq:AdSoff} under the supersymmetric constraint $l g_I W^I = \frac32$,
\be
\label{eq:AdSon}
	\frac{\partial I_{AdS_5}^\text{off-shell}}{\partial W^I} \Big|_{W = \overline W} = 0\ , \quad I_{AdS_5}^\text{on-shell} := I_{AdS_5}^\text{off-shell} (\overline{W} )\ ,
\ee
where $\overline{W}$ is the extremized value of $W$. 
This procedure is the gravitational analog of $a$-maximization~\cite{Anselmi:1997am,Intriligator:2003jj,Kutasov:2003iy}, which in field theory determines the exact superconformal R-symmetry by maximizing the trial $a$-anomaly over all possible R-symmetry mixings. This principle is crucial in distinguishing genuine SCFTs from generic supersymmetric field theories, where the R-symmetry is not fixed dynamically. In this holographic context, we thus interpret the off-shell AdS$_5$ action as the trial $a$-anomaly, while the scalar equations of motion play the role of the $a$-maximization conditions.  Moreover, the constraint appears as the field equation of a certain auxiliary field. Implementing $a$-maximization using off-shell supergravity allows us to go beyond the $N \rightarrow \infty$ limit, thus differing from the literature where discussions were mainly based on the on-shell two-derivative theory~\footnote{Here our perspective is off-shell, thus not directly comparable with the interpretation in \cite{Tachikawa:2005tq,Szepietowski:2012tb}. One can rather view it as complementary to the off-shell geometric perspective of $a$-maximization in terms of the internal Sasakian volume, developed in \cite{Martelli:2005tp,Butti:2005vn,Martelli:2006yb}. See also related results on supergravity localization, \cite{BenettiGenolini:2023kxp,Cassani:2024kjn,Colombo:2025ihp,BenettiGenolini:2025icr}, that would be interesting to revisit in the superconformal formalism.}.

\section{4d anomaly polynomial and (holographic) $a$-maximization}
\label{sec:fieldth}
\vspace{-3mm}
To set the stage for our supergravity analysis, we briefly review key aspects of the holographically dual field theory, following the conventions of~\cite{Cassani:2024tvk}. We consider a 4d $\cN=1$ SCFT with $(n+1)$ global $U(1)$ currents generated by $Q_I$. Without imposing any normalization, their commutation relations with a supercharge $\cQ$ are~\footnote{We leave the details of the spinorial representation of $\cQ$ unspecified, except for its $R$-charge, as they play no role here. See~\cite{Cassani:2024tvk} and references therein for details.}
\be
\label{eq:gencommutator}
[ Q_I, \cQ ] = - r_I\, \cQ\ ,
\ee
where $r_I$ are rational constants. This is not the standard superconformal form, where one distinguishes the $U(1)_R$ generator $R$ from the remaining flavor symmetries $F_i$,
\be
\label{eq:Rcommutator}
[ R, \cQ ] = - \cQ\ , \qquad [ F_i, \cQ ] = 0 \ ,
\ee
with $\cQ$ normalized conventionally. The generators $R$ and $F_i$ are therefore specific linear combinations of the original $Q_I$. We parametrize the $R$-symmetry through a \emph{trial} combination
\be
\label{eq:r-sym}
R^\text{tr} (s) := s^I\, Q_I\ , \qquad r_I\, s^I = 1\ ,
\ee
with the constraint on the $s^I$ following from \eqref{eq:Rcommutator}.

We define the cubic and linear ’t~Hooft anomaly coefficients for the global charges $Q_I$ as
\be
\cA_{IJK} := {\rm Tr}\, (Q_I Q_J Q_K)\ , \qquad \cA_I := {\rm Tr}\, Q_I\ ,
\ee
which for many holographic theories of D3- and wrapped M5-branes are known \emph{exactly} in the gauge group rank $N$, see e.g.\ \cite{Benvenuti:2004dy,Bah:2012dg}. 
From these, one constructs the anomaly polynomial
\be
\label{eq:Apolynomial}
\cA_{4d} = \frac{\cA_{IJK} }{6}\, c_1(\frak{f}^I) \wedge c_1(\frak{f}^J) \wedge c_1(\frak{f}^K) - \frac{\cA_I}{24}\, c_1(\frak{f}^I) \wedge p_1 (T \cY_6)\ ,
\ee
which encodes the variation of the partition function under background gauge transformations of the global symmetries, with field strengths $\frak{f}^I := {\rm d} \frak{a}^I$. Here $\cA_{4d}$ is a six-form defined on an extension $\cY_6$ of the four-dimensional manifold $\cM_4 = \partial$AdS$_5$. In local coordinates,
\be
c_1 (\frak{f}^I) = \frac{\frak{f}^I}{2 \pi}\ , \qquad p_1 (T \cY_6) = \frac1{8 \pi^2}\, \cR_{ab} \wedge \cR^{ab}\ ,
\ee
where $\cR_{ab}$ is the Riemann curvature two-form on $\cY_6$.

When the $R$-symmetry is not manifest, it can be uniquely determined by maximizing the \emph{trial} $a$-anomaly coefficient,
\be
a^\text{tr} (s) := \frac3{32}\, (3\, \cA_{I J K} s^I s^J s^K - \cA_I s^I)\ ,
\ee
whose extremization, subject to \eqref{eq:r-sym}, yields
\be
\frac{\partial a^\text{tr}}{\partial s^I} \Big|_{s = \mathbf{s}} = 0\ , \quad a = a^\text{tr} (\mathbf{s})\ , \quad R = R^\text{tr} (\mathbf{s})\ .
\ee

This procedure, known as $a$-maximization \cite{Anselmi:1997am,Intriligator:2003jj,Kutasov:2003iy}, is the exact field-theoretic counterpart of the gravitational extremization principle~\eqref{eq:AdSon}. The correspondence follows from the holographic relation at the conformal boundary,
\be
{\rm d} \cL_\text{CS} (v) = \cA_{4d} (\frak{a})\ ,
\ee
where we have implicitly extended the l.h.s.\ on $\cY_6$. The holographic dictionary dictates proportionality between the bulk gauge fields and boundary global symmetries, $l^{-1}\, v^I = \frak{a}^I$, as well as $l\, g_I = r_I$, see \cite{Cassani:2024tvk,Klare:2012gn,Cassani:2012ri}, with the latter equality fixed by matching with the R-symmetry charge of the bulk Killing spinor (see next section). This yields the holographic identification between the CS and anomaly coefficients,
\be
k_{IJK} = \frac{1}{2\, \pi^3\, l^3}\, \cA_{IJK} \ , \qquad k_I = \frac1{48\, \pi^3\, l}\, \cA_I\ .
\ee
It was also shown in~\cite{Cassani:2024tvk}, to first order in the four-derivative expansion, that the on-shell action of global AdS$_5$ satisfies
\be
\label{eq:on}
I_{AdS_5}^\text{on-shell} = \frac{3\, \beta}{2\, \pi\, l}\, a\ ,
\ee
where $l$ is the AdS radius and $\beta$ the Euclidean time period, using minimal holographic renormalization. Our analysis extends this result to the exact \emph{trial} $a$,
\be
\label{eq:off}
I_{AdS_5}^\text{off-shell} (W) = \frac{3\, \beta}{2\, \pi\, l}\, a^\text{tr} (s)\ ,
\ee
where the parameters $s^I$ are proportional to the vacuum expectation values of the vector multiplet scalars $W^I$ in the AdS background, $2\, W^I =3\, s^I$, such that the gravitational constraint $l g_I W^I = \frac32$ maps to~\eqref{eq:r-sym}.

\vspace{-5mm}
\section{Multiplets and invariants}
\label{sec:invariants}
\vspace{-3mm}
We now introduce 5d $\cN=1$ superconformal invariants, focusing on a novel class of actions, termed “off-diagonal” in~\cite{Ozkan:2016csy}. To keep the discussion concise, we present only the primary fields of the composite multiplets that describe these invariants. The corresponding descendant components, along with their full superspace formulation, will appear in~\cite{Hristov:2026}.

\paragraph{\bf Fundamental multiplets.}
In five dimensional off-shell supergravity, the standard Weyl multiplet~\cite{Bergshoeff:2001hc} gauges the superconformal algebra $\rm F^2(4)$ and contains $32+32$ off-shell components. The independent gauge fields are the vielbein $e_\m{}^\ha$, gravitino $\psi_\m{}_\hal^i$, ${\rm SU(2)}_R$ gauge field $\phi_\m{}^{ij}$, and dilatation gauge field $b_\m$. The spin connection $\omega_\m{}^{\ha\hb}$, $S$-supersymmetry connection $\phi_\m{}_\hal^i$, and special conformal connection $\mathfrak{f}_\m{}^\ha$ are composite, determined by standard curvature constraints. The multiplet further includes the matter fields: a real antisymmetric tensor $W_{\ha\hb}$, a fermion $\chi_\hal^i$, and a real auxiliary scalar $D$.

An Abelian vector multiplet, denoted henceforth by $W$, consists of a
scalar field $W$~\footnote{By a standard abuse of notation, we denote the complete multiplet in the same way as its lowest component. The distinction should be evident from the context in a given discussion or a formula.}, a gaugino $\l_{\a}^{i}$, a gauge vector $v_{a}$
with covariant field strength $F_{a b}$, and a SU(2) triplet
of auxiliary fields $X^{ij}$. 
The linear multiplet, denoted by $G$, contains an SU(2) triplet of scalars $G^{ij}$, a fermion $\varphi_{\a}^{i}$,  a vector $H^{a}$, Hodge dual of a covariant four-form field strength, and an auxiliary scalar $F$.

\paragraph{\bf The BF action.} In the superconformal tensor calculus, the BF action principle serves as a cornerstone for constructing general supergravity–matter couplings (see~\cite{Butter:2014xxa,Kugo:2000af,Fujita:2001kv,Kugo:2002vc,Bergshoeff:2001hc,Bergshoeff:2002qk,Bergshoeff:2004kh} for the 5d case). This  relies on forming a suitable product between a linear multiplet $(G)$ and an Abelian vector multiplet $(W)$, $S_{\rm BF} := \int {\rm d}^5 x\, e\, \cL_{\rm BF}$,
	\be
	\cL_{\rm BF} (W, G) :=  - \left( W F  + X_{i j} G^{i j} + v_{{a}} H^{{a}} \right)~,
	\label{BF-Scomp}
	\ee
where we neglected the fermions.

Given $(n+1)$ Abelian vector multiplets $W^I$, one can construct $(n+1)$ composite linear multiplets $\underline{G}_I$~\cite{Butter:2014xxa} whose lowest components are 
\be \label{O2composite-N-001}
\underline{G}_{I}^{ij}
= C_{IJK} ( \,2 W^J X^{ij K}
- \ri \l^{\a J\,(i } \l_{\a}^{j) K} )\ ,
\ee
with symmetric constants $C_{IJK}=C_{(IJK)}$. The remaining components of the multiplet, $\underline{\varphi}_{I}^{\a i}$, $\underline{H}_{I}^{a}$ and $\underline{F}_{I}$, are obtained by taking successive $Q$-supersymmetry variations of $\underline{G}_{I}^{ij}$.  Conversely, given a linear multiplet, one can construct a composite vector multiplet with the scalar field
\be \label{comp-W-000-component}
\underline{W} = \frac{1}{4}F G^{-1} - \frac{\ri}{8} G_{i j} \varphi^{i \a} \varphi^{j}_{\a} G^{-3}\ ,
\ee
with $G := \sqrt{\hf G^{ij} G_{ij}}\ne0$. The remaining components of the multiplet, $\underline{\lambda}_{\a}^{i}$, $\underline{F}_{a b}$ and $\underline{X}^{ij}$, are obtained by taking successive $Q$-supersymmetry variations of $\underline{W}$. The explicit construction of the two-derivative supergravity theory, including the substitution of the composite multiplets into the BF action is presented in section A of the Supplemental Material. Here we simply note that, as a result of substituting the composite multiplets \eqref{O2composite-N-001} into the BF action \eqref{BF-Scomp}, the coefficient of the Chern–Simons term is found to be proportional to $C_{IJK}$ of Eq.\,\eqref{O2composite-N-001}. 

All off-shell superconformal invariants considered in our analysis—including the two- and four-derivative terms—are listed in Table~\ref{tab:solutions} together with their associated references. For the purposes of this letter, we briefly outline the construction of the curvature-squared and off-diagonal invariants coupled to $(n+1)$ vector multiplets. 

\paragraph{\bf The three curvature-square invariants.} 
Here we present the primary fields corresponding to the composite multiplets that generate the three curvature-squared invariants—namely, the Weyl-squared, Ricci-squared, and scalar curvature-squared terms:
\bea
G^{i j}_{\textrm{Weyl}^2} &:=& - \frac{\ri}{2}   W^{{\alpha} {\beta} {\gamma} i} W_{{\alpha} {\beta} {\gamma}}\,^{j}-  W^{a b} \phi_{a b}\,^{i j} + \frac{256 \ri}{3}   \chi^{{\alpha} i} \chi^{j}_{{\alpha}}~, \nonumber \label{H-Weyl-0} \\  
G^{i j}_{\log} &:=& - \frac{3 \ri}{40} \Delta^{i j k l}\de_{k l} \log{(\sigma_I W^I)} ~, \nonumber \\
G^{ij}_{\cR^2} &:=& \underline{G}^{ij}[\underline{W}] =
2\underline{W} \underline{X}^{ij}
-\ri \underline{\l}^{\a(i}\underline{\l}_{\a}^{j)} ~,
\label{composite-diagonal-G}
\eea
and corresponding actions $\cL_{\rm BF} (\alpha_I W^I, G_{\textrm{Weyl}^2})$, $\cL_{\rm BF} (\beta_I W^I, G_{\log})$, $\cL_{\rm BF} (\gamma_I W^I, G_{\cR^2})$, respectively. Note that the definition of the matter-coupled $\log$-invariant, for which we refer the reader to \cite{Butter:2014xxa,Gold:2023ykx} for its technical construction, allows for additional arbitrary coefficients $\sigma_I$ in the action, which drop out of the final results.  The explicit off-shell Chern–Simons terms corresponding to the four-derivative invariants are listed in section~B of the Supplemental Material.

\paragraph{\bf Off-diagonal invariants.} 
We now turn to a wider class of \emph{off-diagonal} invariants, obtained by coupling the composite vector multiplets $\underline{W}$ and $W_{I}^{F^4}$ (see below) to the composite linear multiplets $\underline{G}_I$, $G_{\rm Weyl}$, $G_{\rm log}$, and $G_{\cR^2}$. These interactions generate terms beyond the curvature-squared invariants associated with \eqref{composite-diagonal-G}. While a subset was identified in~\cite{Ozkan:2016csy}, here we provide a classification of such invariants relevant to higher-derivative deformations in supergravity. The number of derivatives that these actions carry off-shell is most directly seen by examining the BF-action, \eqref{BF-Scomp}, together with the following tables for composite multiplet components.
    
    \vspace{-0.21cm}
\begin{table}[ht]
\centering
\begin{minipage}{0.23\textwidth}
\centering
\setlength{\tabcolsep}{6pt}
\begin{tabular}{ c || c | c | c | } \hline
\multicolumn{1}{|c||}{\textbf{linear}} & $G^{ij}$ & $H^a$  & $F$  \\ \hline \hline
\multicolumn{1}{|c||}{$\underline{G}_I$}  & $0 \partial$ & $2 \partial$ & $2 \partial$ \\ \hline
\multicolumn{1}{|c||}{\text{Weyl}$^2$} & $2 \partial$ & $4 \partial$ & $4 \partial$ \\ \hline
\multicolumn{1}{|c||}{$\log$} & $2 \partial$ & $4 \partial$ & $4 \partial$ \\ \hline
\multicolumn{1}{|c||}{$\cR^2$} & $2 \partial$ & $4 \partial$ & $4 \partial$ \\ \hline
\end{tabular}
\end{minipage}
\hfill
\begin{minipage}{0.23\textwidth}
\centering
\setlength{\tabcolsep}{6pt}
\begin{tabular}{ c || c | c | c | } \hline
\multicolumn{1}{|c||}{\textbf{vector}} & $W$ & $v_a$  & $X^{ij}$  \\ \hline \hline
\multicolumn{1}{|c||}{$\underline{W}$}  & $0 \partial$ & $0 \partial$ & $2 \partial$ \\ \hline
\multicolumn{1}{|c||}{$F^4$} & $2 \partial$ & $2 \partial$ & $4 \partial$ \\ \hline
\end{tabular}
\end{minipage}
    \vspace{-0.1cm}
\caption{The derivative structure (over fundamental fields) of the bosonic components of the composite linear and vector multiplets used here. Combining each row on the left with a row on the right in a BF action produces a different off-diagonal invariant.}
\label{tab:linear}
\end{table}
    \vspace{-0.15cm}

A two-derivative~\footnote{Note that classifying superconformal invariants by their derivative order is not unique. Because gauged supergravity contains a natural length scale, $l$, one may instead classify invariants according to their scaling in $l$, which can be read off from Table~\ref{tab:solutions} and Eq.~\eqref{eq:AdSoff}. From this perspective, the $\cR F$ invariant exhibits a scaling behavior distinct from that of the standard two-derivative action.} supersymmetric completion of the $\cR F$ term arises from coupling the composite $\underline{W}$ to the linear multiplet $\underline{G}_{I}$ within the BF action~\eqref{BF-Scomp},
\be
    \cL_{\cR F} :=  \cL_{\rm BF} (\underline{W}, a^I \underline{G}_I) 
    ~,
\ee
where, in~\eqref{O2composite-N-001}, $C_{IJK}$ are replaced by $d_{IJK}$ (another set of arbitrary coefficients).

A four-derivative supersymmetric completion of the $\cR F^3$ term arises  by coupling the composite vector $\underline{W}$ and linear $G_{\cR^2}$ multiplets within the BF action~\eqref{BF-Scomp},
\be
    \cL_{\cR F^3} := \eta\, \cL_{\rm BF} (\underline{W}, G_{\cR^2}) 
    ~.
\ee

In analogy to these constructions, we are able to generate a couple of new superconformal invariants, replacing $G_{\cR^2}$ above by $G_{\log}$ or $G_{\textrm{Weyl}^2}$, which will be discussed at length in \cite{Hristov:2026}. Here we finish with a supersymmetric completion of the $F^4$ term arising from a composite vector multiplet with the primary scalar field,~\cite{Ozkan:2016csy}
\be
\begin{split}
     W_{I}^{F^4} &:= \frac{1}{4 G} \underline{F}_{I} -  \frac{1}{8 G^3} \Big( G_{i j} \underline{G}_{I}^{ij} {F} + 2 \ri G_{i j} \varphi^{\a i} \underline{\varphi}_{\a I}^{j}\\ &+ \ri \underline{G}_{I}^{i j} \varphi^{\a}_{i} \varphi_{\a j} \Big)+ \frac{3 \ri}{16 G^5}G_{ij} G_{kl} \underline{G}_{I}^{k l} \varphi^{\a i} \varphi_{\a}^{j} ~, \label{composite-F^4}
\end{split}
\ee
coupled to another set of composites $\underline{G}_{J}$, 
\be
    \cL_{F^4} := a^{IJ}\, \cL_{\rm BF} (W^{F^4}_I, \underline{G}_J)\ ,
\ee
resulting in a four-derivative invariant. Combining the $F^4$ multiplet with the other linear multiplets results in new invariants, also to be discussed in \cite{Hristov:2026}.

\paragraph{\bf On-shell Chern-Simons terms.} 
 The explicit off-shell Chern–Simons terms for the four-derivative invariants are collected in section~B of the Supplemental Material. Unlike the two-derivative case, the Chern–Simons term at the four-derivative level often appears through the auxiliary field $\phi_{a}^{ij}$. To extract the correct Chern–Simons coefficient, one must substitute $\phi_{a}^{ij}$ using the leading two-derivative equation of motion of $\phi^{a}_{ij}$ and $H^a$,
\be
   \phi_{a}^{ij} = - g_{I} v_{a}^{I} \delta^{ij}~, \qquad H_{a} = 0~.
   \label{phi-H-1}
\ee
The resulting coefficients of the Chern–Simons terms for all invariants are summarized in Table~\ref{tab:solutions}. Importantly, Eq.\,\eqref{phi-H-1} also gives the Killing spinor R-symmetry charge that we used previously in the holographic dictionary, via the covariant derivative $\cD_\mu \epsilon^i = \partial_\mu \epsilon^i + \phi_{\mu}^{ij} \epsilon_j + \cdots $. 

\vspace{-5mm}
\section{Partially off-shell AdS$_5$ action}
\label{sec:action}
\vspace{-3mm}
We now describe the maximally supersymmetric AdS$_5$ background, which provides the setting for evaluating the invariants constructed above. The final results for each invariant, summarized in~\eqref{eq:AdSoff} using the table of Chern–Simons coefficients, have already been anticipated. We therefore omit intermediate details, which will be presented in~\cite{Hristov:2026} comprehensively. Here we focus on the conditions satisfied by the relevant multiplet components under maximal supersymmetry, without imposing any equations of motion. 

Since the $U(1)_R$ symmetry—gauged by a linear combination of the Abelian vectors—embeds into the original $SU(2)_R$ symmetry along an arbitrary direction, we introduce a unit tensor $s^{ij}$ satisfying $s^2 = \tfrac12\, s^{ij} s_{ij} = 1$. We will choose $s^{ij}=\delta^{ij}$. Maximal supersymmetry then constrains all $SU(2)_R$ tensors in the theory to align with this direction, i.e.\ $X^{ij}$ and $G^{ij}$ are proportional to $\delta^{ij}$. For the metric, we take AdS$_5$ in the global coordinates,
\be
\label{eq:ads5metricins1s3coordinates}
	{\rm d} s^2_5 = -  \left( 1 + \frac{r^2}{l^2} \right)\, {\rm d} t^2 + \frac{{\rm d} r^2}{1 + \frac{r^2}{l^2}} + r^2 s^2_{S^3}\ ,
\ee
where $ds^2_{S^3}$ is the metric on $S^3$ with unit radius, such that the Ricci scalar is given by $\cR = 20\, l^{-2}$ ~\footnote{In the present convention for curvature, AdS$_5$ has a constant positive curvature.}. We further find that all other Weyl multiplet components vanish, notably including the auxiliary scalar field, $D=0$. For the linear multiplet,  maximal supersymmetry, together with the standard gauge fixing choice $G=1$, sets
\be
	G^{ij}=
    \, \d^{ij} ~,~~~ \varphi_\b^i=0~,~~~H_a=0~,~~~F={\rm const}~.
\ee
Similarly, for a vector multiplet we find
\be
	W^I={\rm const} ~,~~~ \l_\a^{i I}=0 ~,~~~ F^I_{ab}=0~, ~~~ X^{ij I}=\frac1{6}F\, W^I\d^{ij}~.
\ee
 All remaining multiplets, being composite, follow by an insertion of the identities above in their definition.

In order to truly evaluate the resulting actions, we need the additional input of the field equation of the auxiliary field $F$, which only gets a two-derivative contribution:
\be
	F = -\frac6{l} = -4\, g_I W^I\ ,
\ee
giving an important constraint on the corresponding linear combination of vector multiplet scalars, since the AdS scale $l$ needs to be taken into account in the holographic dictionary. This is the origin of the constrained extremization in \eqref{eq:AdSoff}, where the field $F$ can be regarded as the holographic dual of the  Lagrange multiplier enforcing \eqref{eq:r-sym}.

With the above identifications, we have been able to explicitly evaluate the AdS$_5$ background on each of the Lagrangians we constructed, see more details in the Supplemental Material. In turn we need to discuss the holographic renormalization scheme that renders the corresponding actions finite.
Importantly, the choice of boundary in this case, $\mathbb{R} \times$S$^3$, does not allow for any finite counterterms or logarithmic divergences, rendering our results unambiguous. The holographically renormalized action can be simply computed in the following way,
\be
	I_M = \lim_{r \rightarrow \infty}\, S_\text{sugra} (M) +\sigma_\text{ct}\,  S_\text{GH} (\partial M) +\sigma_\text{ct}\, S_\text{ct} (\partial M)\ ,
\ee
with $S_\text{sugra} (M)$ the original supergravity action with asymptotically AdS$_5$ background as in \eqref{eq:ads5metricins1s3coordinates}. The Gibbons-Hawking action is given by
\be
\label{eq:25}
	S_\text{GH} (\partial M) = - \int_{\partial M} d^4 x\, \sqrt{-h}\, K\ ,
\ee
where $h_{ij}$ is the induced metric on the boundary, and $K = h^{ij}\, K_{ij}$ is the trace of the extrinsic curvature. The final counterterm is given by
\be
\label{eq:26}
	 S_\text{ct} (\partial M) = \int_{\partial M} d^4 x\, \sqrt{-h}\, \left( - \frac3{l} + \frac{l\, R(h)}4 \right)\ ,
\ee
with $R(h)$ the scalar curvature of the induced metric $h$, which renders the full answer finite. 

Note that we have allowed an arbitrary overall normalization of the original action. The constant $\sigma_\text{ct}$ should then be chosen accordingly, such that the full answer is indeed finite. We consequently recover the partially off-shell results as advertised in the introduction.

\vspace{-5mm}
\section{Outlook}
\label{sec:outlook}
\vspace{-3mm}
Building on the relation between the anomaly polynomial and the superconformal invariants established above, we conclude with an observation that merits further exploration. For holographic 4d SCFTs at large $N$, one expects that all supersymmetric partition functions (or large-$N$ protected observables) are completely determined by the ’t Hooft anomaly coefficients of the R-symmetry and flavor symmetries. Holographically, this suggests that the (partially) off-shell action for any supersymmetric asymptotically locally AdS$_5$ background is entirely determined by the Chern–Simons coefficients listed in Table~I. Moreover, while higher-derivative invariants involving six or more derivatives can, in principle, be constructed within the present formalism (see~\cite{Butter:2014xxa,Hristov:2026}), dimensional analysis implies that they cannot generate additional on-shell Chern–Simons terms beyond those in~\eqref{eq:on-shellCS}. We thus expect all such higher-order invariants, when evaluated on asymptotically AdS$_5$ BPS backgrounds, to reduce to linear combinations of the two-derivative and Weyl$^2$ actions, which already exhaust the possible freedom in the coefficients $k_{IJK}$ and $k_I$. This conclusion applies to all other invariants considered here, and will be extended to include a more complete set of invariants and building blocks in \cite{Hristov:2026}. By employing the formalism of \cite{Kuzenko:2007cj,Butter:2014xxa}, it would naturally be very interesting to include in our analysis gauged hypermultiplets for a complete off-shell description.

\vspace{-5mm}
\section*{Acknowledgements}
\vspace{-3mm}
We thank Robert Saskowski for discussions. K.~H. is supported in part by the Bulgarian NSF grant KP-06-N88/1. Y.~P. is supported by the National Key R\&D Program No. 2022YFE0134300 and the National Natural Science Foundation of China (NSFC) under Grant No. 12575076, No.12247103.
G. T.-M. have been supported by the Australian Research Council
(ARC) Future Fellowship FT180100353, ARC Discovery Project DP240101409, the Capacity Building Package at the University of Queensland, and a faculty start-up funding of UQ’s School of Mathematics and Physics. 
We thank support during the MATRIX Program “New Deformations of Quantum Field and Gravity Theories,” (Creswick, 22 Jan – 2 Feb 2024), where this collaboration was initiated.

\bibliographystyle{apsrev4-2}
\bibliography{refs.bib}

\clearpage

\appendix{}

\onecolumngrid

\setcounter{section}{1}
 \setcounter{equation}{0}
 
\begin{center}\textbf{SUPPLEMENTAL MATERIAL\\(APPENDICES)}\end{center}

\vspace{-0.8 cm}
\section{\large 
\textbf{
A.  Two derivative theory}}\label{app:twoderiv}

In components, substituting the composite multiplets \eqref{O2composite-N-001}, and \eqref{comp-W-000-component} in the BF action \eqref{BF-Scomp} leads to the action for $(n+1)$ Abelian vector multiples (including the compensator) and linear multiplet compensator. The bosonic part of these actions takes the form $\cL_{2 \partial} := \cL_{\rm VM} + \cL_{\rm L} + \cL_{\rm g}$:
\bsubeq
\bea
\cL_{\rm VM}
&:=& \cL_{\rm BF} (W^I, \underline{G}_I) =  C_{IJK}\Big( \frac{1}{8} \ve_{\ha \hb \hc \hd \he}v^{\ha I}F^{\hb \hc J} F^{\hd \he K}
- \frac{1}{8} W^{I} W^{J} W^{K} {\cR}
+ \frac{3}{2} W^{I} ({\cD}^{\ha} W^{J}) {\cD}_{\ha} W^{K} 
\\
&&
- \frac{3}{4} W^{I} X^{ij J} X_{ij}^{K}
+ \frac{3}{4} W^{I} F^{\ha \hb J} F_{\ha \hb}^{K}
+ \frac{9}{4} W^{I} W^{J} W^{\ha \hb} F_{\ha \hb}^{K}
+ \frac{39}{32} W^{I} W^{J} W^{K} W^{\ha \hb}W_{\ha \hb} 
+ 4 W^{I} W^{J} W^{K} D
\Big)~,
\label{bosonic-VM-swm} \non \\
\label{eq:TensorComp}
\cL_{\rm L} 
&:=& \cL_{\rm BF} (\underline{W}, G) =
- \frac{3}{8} \cR 
-4 D 
 - \frac{1}{8} F^2
- \frac{3}{32} W^{ a b} W_{ a b} 
+ \phi^{\prime}_{a ij} \phi^{\prime a ij}
- \frac{1}{8}  H^a H_a
- \frac{1}{2} H^a \phi_a
~, \\
\cL_{\rm g} &:=& \cL_{\rm BF} (g_I W^I, G) = - g_{I} W^{I} F - g_{I} v^{I}_{{a}} H^{{a}}  - g_{I} X^{I}_{i j} \delta^{i j}  \label{bosonic-gauged-term}~.
\eea
\esubeq
Here we have decomposed $\phi_{a}^{i j}$ into traceless and trace parts $\phi_{a}^{i j} = \phi_{ a}^{\prime i j} + \tfrac{1}{2} \delta^{i j} \phi_{a}$ and used $G^{ij}=\delta^{ij}$ as part of the gauge fixing. The coefficient of the Chern–Simons term follows directly from this action. From Eq.~\eqref{eq:on-shellCS}, we can then identify the cubic and linear anomaly coefficients as $k_{IJK}^{2\partial} = 3\, C_{IJK}, \, k_I^{2\partial} = 0$.

Based on the above theory, the equations of motion for the auxiliary fields are
\bea
    \delta X_{ij}^{I}: && C_{IJK} W^{J} X^{ij\, K} = -\tfrac{2}{3}\, g_{I} \delta^{ij}\ , 
    \nn\\
  \delta   F: && F = -4\, g_{I} W^{I}\ ,  
  \nn\\
  \delta  D: && C_{IJK} W^{I} W^{J} W^{K} = 1 \ , 
  \nn\\
  \delta W^{ab}: && W_{ab}  = -C_{IJK} W^{I} W^{J} F_{ab}^{K}\ ,  
  \nn\\
 \delta \phi_a: && H^{a} = 0\ , 
 \nn\\
 \delta \phi^{\prime a}_{ij}: && \phi^{\prime\, ij}_{a} = 0\ ,
 \nn\\
\delta    H^{a}: 
   && \phi_{a} = -2 g_{I} v_{a}^{I} - \tfrac{1}{2} H_{a}\ .
  \label{EOM-2derivative-n-VM} 
\eea  
In order to write the actions above, we used that the composite vector field $\underline{v}_{a}$ descending from \eqref{comp-W-000-component} is given by
\be
\underline{v}_{a}=\tfrac12 \phi_a=-g_Iv^I_a\ , \quad \Rightarrow \quad \underline{F}_{ab} = - g_I F^I_{ab}\ ,
\ee
where we used \eqref{EOM-2derivative-n-VM}.

\vspace{0.15cm}
\begin{center}{\large 
\textbf{
B.  Chern-Simons and scalar terms }}
\end{center}
\vspace{0.15cm}
Here we summarize the Chern-Simons terms along with the scalar terms that contribute to the action on the AdS$_5$ background for the various superconformal invariants,
\bsubeq
\bea
 \cL_{\cR F} &=&  \frac{1}{2}  a^K d_{IJK}  \,   \ve_{{a} {b} {c} {d} {e}} \underline{v}^{{a}} F^I{}^{{b} {c}} F^{J}{}^{{d} {e}} - a^I
\underline{W} \underline{F}_{I}
- a^I  \underline{X}_{ij} \underline{G}_{I}^{ij}+\cdots, 
\\
\cL_{\text{Weyl}^2} 
&=& -\frac{\alpha_I}{8}\,\varepsilon^{a b c d e}\,v^{I}_{a}\,\cR_{b c}\,^{f g}\,\cR_{d e f g}
+ \frac{\alpha_I}{6}\,\varepsilon^{a b c d e}\,v^{I}_{a}\,\Phi_{b c}\,^{i j}\,\Phi_{d e\, i j} -\alpha_I 
W^I F_{{\rm Weyl}^2}
-\alpha_I  X^{I}_{ij} G^{ij}_{{\rm Weyl}^2}+\cdots, 
\\
 \cL_{\log} &=&   - \frac{1}{12} \, \beta_I \, \ve^{{a} {b} {c} {d} {e}} v^{I}_{{a}} \Phi_{{b} {c}}\,^{i j} \Phi_{{d} {e} i j} -\beta_I 
W^I F_{\rm log}
-\beta_I  X^{I}_{ij} G^{ij}_{\rm log}+\cdots, 
\\
\cL_{\cR^2} &=& \frac{1}{2}\, \gamma_I\,  \ve_{{a} {b} {c} {d} {e}} v^{I {a}} \underline{F}^{{b} {c}} \underline{F}^{{d} {e}} -\gamma_I 
W^I F_{\cR^2}
-\gamma_I  X^{I}_{ij} G^{ij}_{\cR^2}+\cdots\ , 
\\
 \cL_{\cR F^3} &=& \frac{1}{2}  \eta  \,      \ve_{{a} {b} {c} {d} {e}} \underline{v}^{{a}} \underline{F}^{{b} {c}} \underline{F}^{{d} {e}} - \eta
\underline{W} F_{\cR^2}
- \eta  \underline{X}_{ij} G^{ij}_{\cR^2} +\cdots, 
\\
\cL_{F^4} &=& 
- a_{IJ} {W}^{I}_{ F^4} \underline{F}^{J}
- a_{IJ}  {X}^{I}_{ F^4}{}_{ij} \underline{G}^{J ij}+\cdots~.
\eea
\esubeq
Here the constants $\alpha_I, \beta_I, \gamma_I, \eta, a^{I}, d_{IJK}, a^{IJ}$ are the various coupling constants of the corresponding supersymmetric actions, and $\Phi_{a b}{}^{ij} := 2\,e_\ha{}^\hm e_\hb{}^\hn \big( \pa_{[\hm} \phi_{\hn]}{}^{ij}
	+ \phi_{[\hm}{}^{k (i} {\phi}_{\hn]}{}^{j)}{}_{k }\big)$. Note that, in order to find the on-shell CS actions in the form of \eqref{eq:on-shellCS}, it is enough to use the two-derivative equations of motion given above. The reason is that further corrections from higher-derivatives can in turn be written in a covariant way due to the additional derivatives and do not constitute genuine CS terms.

The underlined fields are the composite fields of the two-derivative theory, given in \eqref{O2composite-N-001} and \eqref{comp-W-000-component}. When evaluated on a supersymmetric $AdS_5$ background, they take the form
\bsubeq
\begin{align}
&\underline{G}_I:     && \underline{G}_I =2 \, C_{IJK}  W^J X^{K} ~,~~~
\underline{F}_I = C_{IJK} \left( 2X^{J}X^{K} + \frac{10}{l^2} W^J W^K - 16 D W^J W^K \right) ~, \\
&\underline{W}:  && \underline{W} = \frac{1}{4} F   ~,
~~~
\underline{X} =  \left(\frac{15}{4 l^2} +2 D -\frac{1}{16} F^2 \right) ~,
\end{align}
\esubeq
where maximal supersymmetry constrains $\tilde{X}^{ij} = \delta^{ij} \tilde{X}$ and $\tilde{G}^{ij} = \delta^{ij} \tilde{G}$, where $\tilde{X}^{ij}$ and $\tilde{G}^{ij}$ collectively represent the various $SU(2)_R$ tensors in the theory.

The following are the composite fields of the higher-derivative invariants, which when evaluated on the $AdS_5$ background, and retaining terms only up to linear order in $D$, take the form
\bsubeq
\begin{align}
 & {\rm Weyl}^2:   && G_{\textrm{Weyl}^2} = 0
    \label{H-Weyl-01}~, ~~~
 F_{\textrm{Weyl}^2} = 0 ~, \\
 & {\rm log}: && 
    G_{\log}=    -6  (\sigma_I W^I)^{-1}\sigma_J X^JD 
         -\frac{20}{16 l^2}  (\sigma_I W^I)^{-1}\sigma_J X^J
    - \frac{1}{4}(\sigma_I W^I)^{-3} (\sigma_J X^J)^3~, \\
   &&& F_{\log} 
    =
    -\frac{85}{8 l^4}
    +\frac{5}{4 l^2} (\sigma_I W^I)^{-2} (\sigma_J X^J)^2
    +\frac{3}{8}(\sigma_I W^I)^{-4} (\sigma_J X^J)^4
    +\frac{10}{l^2} D
    + 6(\sigma_I W^I)^{-2}  (\sigma_J X^J)^2 D~, \\
  & \cR^2: && G_{\cR^2} =
     \frac{15}{8 l^2} F  
    -\frac{1}{32}  F^3 
    +  F D~, \\
   &&& F_{\cR^2} =
    \frac{225}{8 l^4}
    -\frac{5}{16 l^2} F^2 
    +\frac{1}{128}  F^4
    -\frac{3}{2}   F^2 D 
    +\frac{30}{l^2} D~, \\
 &  F^4: && W_{I}^{F^4} = \frac{1}{4}{\underline{F}_I} -  \frac{1}{ 4}\, {\underline{G}_{I}} {F} = 0~,\\
   &&& X_{I}^{F^4} = -\frac{1}{8 }\, F \underline{F}_{I} + \frac{1}{8} {\underline{G}_{I}} F^2 = 0~.
\end{align}
\esubeq
We have preserved the linear orders in $D$ as shown above, even though $D$ itself vanishes on the AdS$_5$ backgrounds, in order to derive its equation of motion. The resulting equation can be demonstrated to be consistent with the extremization of the AdS$_5$ action, \eqref{eq:AdSon}.

\end{document}